\documentclass[10pt,conference]{IEEEtran}
\IEEEoverridecommandlockouts
\usepackage{cite}
\usepackage{amsmath,amssymb,amsfonts}
\usepackage{algorithmic}
\usepackage{graphicx}
\usepackage{textcomp}
\usepackage{xcolor}
\usepackage{verbatim}
\usepackage{comment}
\usepackage{color,soul}
\usepackage{multirow}
\usepackage{blindtext}
\usepackage{enumitem}

\usepackage{tikz}

\def\BibTeX{{\rm B\kern-.05em{\sc i\kern-.025em b}\kern-.08em
    T\kern-.1667em\lower.7ex\hbox{E}\kern-.125emX}}

\definecolor{amethyst}{rgb}{0.6, 0.4, 0.8}
\newcommand{\yixue}[1]{{\color{amethyst}(Yixue: #1)}}

\newcommand{\appr}{\textsc{MAOMAO}}
\newcommand{\paloma}{{PALOMA}}

\begin{document}

\title{A Microservice Architecture for \\Online Mobile App Optimization
}

\author{
Yixue Zhao \\
University of Southern California \\
yixue.zhao@usc.edu
\and
Nenad Medvidovic \\
University of Southern California \\
neno@usc.edu
}	

\maketitle

\begin{abstract}
A large number of techniques for analyzing and optimizing mobile apps have emerged in the past decade. However, those techniques' components are notoriously difficult to extract and reuse outside their original tools. This paper introduces \appr, a microservice-based reference architecture for reusing and integrating such components. \appr's twin goals are (1) adoption of available app optimization techniques in practice and (2) improved construction and evaluation of new techniques. The paper uses several existing app optimization techniques to illustrate both the motivation behind \appr~and its potential to fundamentally alter the landscape in this area.
\end{abstract}


\section{introduction}
\label{sec:intro}

Over the past decade, mobile computing devices have become dominant~\cite{wearesocial} and this trend is bound to continue into the foreseeable future. 
New technologies invariably bring challenges that require researchers' attention, 
such as the problems in the mobile domain that affect mobile app performance~\cite{paloma_icse,liu2014characterizing,liu2015automated,wang2015earlybird,zhang2013cachekeeper,yan2012fast,pathak2012keeping}, energy use~\cite{banerjee2014detecting,li16bouquet,banerjee2016automated,liu2014greendroid,rat2018issta,liu2015diagnosing,li2014empirical}, and security ~\cite{lee2017sealant,li2015iccta,li2015apkcombiner,avdiienko2015mining,hammad2018self,arzt2014flowdroid,wei2014amandroid}. 

We have studied the recent work in this domain to identify emerging research trends as well as  problems that remain unaddressed. We have found that the research in the mobile app domain is still at a relatively early stage, and that there is a pronounced gap between research and practice. The majority of existing work still focuses on \textit{identifying} problems, such as detecting performance bugs~\cite{liu2014characterizing,liu2015automated}, security vulnerabilities~\cite{li2015iccta,lee2017sealant}, and energy hotspots~\cite{banerjee2014detecting,li2014empirical}, while techniques for  \textit{solving} them are invariably left to future work. 

As an illustration, 48 papers have dealt with mobile apps in the last five ICSEs (2014--2018). However, only 7 of those papers propose a technique that aims to  optimize an app to address an identified problem. 
Even then, these techniques are usually evaluated on limited numbers of real apps, making their practical applicability unclear. For example, PALOMA~\cite{paloma_icse} was evaluated on 32 apps and Bouquet~\cite{li16bouquet} on only 5. Furthermore, these techniques are hard to adopt in practice because they usually involve non-trivial steps, such as advanced program analysis, that are likely to have a prohibitively steep learning curve for most app developers. 

To identify the reasons behind the dearth of app optimization techniques and their  lack of adoption in practice, we contacted the authors of several techniques and studied published causes of research-industry barriers~\cite{harmanstart,sherif2003barriers,cordy2003comprehending,redwine1985software}. We discovered several common themes. In the research community, (1) it is challenging to find subject apps that fit a given target problem;  (2) app optimization techniques are usually built on top of research tools that have limited documentation and  technical support; (3) it is hard to simulate real-world scenarios in a lab environment.  
From the practitioner's side, (1) it is hard to find  research techniques that solve the exact problems a developer faces; (2) the research techniques are usually not evaluated in large-scale, real scenarios, rendering any claims unconvincing; (3) research techniques usually have limited documentation, making them difficult to adopt by app developers with little-to-no knowledge in a specific research area.

We believe the fundamental problem behind this gap is the lack of a standard protocol to guide the development of research techniques and 
to connect developers and researchers in a way that leverages each side's expertise.
Specifically,  individual app optimization techniques are designed in ad-hoc ways that hinder their reusability and composability. 
To address the problem, we propose a \textbf{M}icroservice \textbf{A}rchitecture for \textbf{O}nline \textbf{M}obile \textbf{A}pp \textbf{O}ptimization (\appr) and a corresponding \textbf{M}icroservice \textbf{R}epository (MR). \appr~is a reference architecture for mobile app optimization techniques that is intended to be comprehensive in scope, but simple enough to be easily extensible. 
 MR is a cloud-based repository to deploy \appr-compatible techniques that connects  researchers  and developers by providing a shared baseline.   


In this paper, Section~\ref{sec:example} summarizes representative existing techniques and describes their (often missed) reuse opportunities. Section~\ref{sec:maomao} introduces  \appr~and discusses how existing techniques can be migrated to it. Section~\ref{sec:vision} elaborates our vision for adopting \appr~in practice. Section~\ref{sec:conclusion} provides concluding thoughts and outlines the future work.
\vspace{-1mm}
\section{Existing Techniques}
\label{sec:example}
\vspace{-1mm}

In this section, we introduce several independently developed techniques that focus on different problems in the mobile domain. 
We highlight each technique's major components to illustrate the potential (and, in practice, often missed) reuse opportunities. We will use these as well as other existing techniques to demonstrate how \appr's architecture can integrate disparate existing solutions (Section~\ref{sec:maomao}) and facilitate their reuse by both developers and researchers (Section~\ref{sec:vision}). 

\textbf{PALOMA}~\cite{paloma_icse}  reduces app latency by prefetching HTTP requests, via four major components: (1) \textit{String Analyzer} identifies suitable HTTP requests for prefetching by interpreting their URL values; (2) \textit{Callback Analyzer} detects the program points to issue prefetching requests; (3) \textit{Instrumenter} uses the above information to produce a prefetching-enabled app; (4) at app runtime, the instrumented app triggers {PALOMA}'s \textit{Proxy} to issue prefetching requests and cache prefetched responses. 

\textbf{IMP}~\cite{higgins2012informed} is a cost-benefit analysis that decides when and how much data to prefetch in an app, via three major components: (1) \textit{API Support} provides ``hints'' on what to prefetch; (2) \textit{Monitor} monitors mobile device's network bandwidth, data usage, and battery status; (3) \textit{Prefetcher}  adapts prefetching strategies based on the ``hints'' and runtime resource usage.

\textbf{Bouquet}~\cite{li16bouquet} bundles HTTP requests to reduce the energy consumption of an app, via three major components: (1) \textit{Detector} of Sequential HTTP Request Sessions (SHRSs), where triggering the first request implies the following requests will also be made; (2) \textit{Bundling Analyzer} generates code to bundle each SHRS; (3) \textit{Proxy} intercepts HTTP requests and runs the bundling code to return corresponding SHRS responses.

Many existing app optimization techniques focus on security. We highlight three representative examples. \textbf{IccTA}'s~\cite{li2015iccta}  \textit{Taint Analyzer}  detects privacy leaks among an app's components. 
\textbf{SEALANT}'s~\cite{lee2017sealant} \textit{Analyzer} also identifies such leaks, while its \textit{Interceptor} manages inter-app interactions to block the leaks. \textbf{ApkCombiner}'s~\cite{li2015apkcombiner} \textit{Combiner} 
compiles multiple apps together to support inter-app privacy leak detection. 


We see a notable reuse opportunity among these techniques. For instance, {PALOMA} and {Bouquet} can reuse {IMP}'s \textit{Monitor} and \textit{Prefetcher} to dynamically adapt their strategies for issuing HTTP requests based on the runtime resource constraints.
{Bouquet}'s \textit{Detector} can reuse {PALOMA}'s \textit{String Analyzer} to interpret the URL values when identifying SHRSs.
{PALOMA} currently targets individual apps, but {ApkCombiner}'s \textit{Combiner} would enable PALOMA to prefetch HTTP requests across  apps. In another scenario, 
{SEALANT}'s \textit{Analyzer} and {IccTA}'s \textit{Taint Analyzer} may be employed in tandem, either to directly compare their results (benefiting researchers) or to leverage their respective strengths (benefiting app developers).

However, reusing and combining existing research techniques is not a simple task in practice: their internal designs may not be properly modularized, their implementations may not be publicly available, and their documentation may be inadequate. 
Reuse and combination of different techniques' capabilities currently tends to require  close communication with the authors of a given technique. 
This  is  time-consuming and unpredictable, resulting in regularly missed reuse opportunities and duplication of work. As a result,  the above  techniques have been successfully used in tandem in only two instances to our knowledge: {SEALANT} uses {IccTA}'s \textit{Taint Analyzer} to evaluate its own \textit{Analyzer}'s accuracy, while {ApkCombiner} reuses {IccTA} to detect inter-app privacy leaks. In the latter case, {ApkCombiner}~\cite{li2015apkcombiner} and {IccTA}~\cite{li2015iccta} share authors, which only further reinforces our point. 
\vspace{-1mm}
\section{\appr}
\label{sec:maomao}
\vspace{-1mm}

We design \appr~based on the existing techniques, such as those highlighted above, and our own experience in the mobile domain. Our aim is to render reusable components at a proper granularity that can, both, serve as a roadmap for future techniques and improve the reusability of  existing techniques. 
In this section, we introduce the design of \appr's architecture (Section~\ref{sec:maomao:design}) and elaborate how existing techniques can be integrated with \appr~ (Section~\ref{sec:maomao:implement}).

\subsection{\appr's Design}
\label{sec:maomao:design}
\vspace{-1mm}

\appr's design is based on the widely-adopted microservice architectural style because (1) it helps to decouple potentially complex functionality into lightweight, ``black-box'' microservices, which are easy to understand and adopt by developers in practice~\cite{villamizar2015evaluating}; (2) existing mobile techniques tend to comprise clearly separable and often reusable components, and the microservice style would make it easier to reuse such components across techniques; (3) the microservice style allows components (i.e. microservices) to be implemented in different programming languages with different technologies, which suits the heterogeneity of the mobile  domain.

\begin{figure}
	\centering
		\includegraphics[width=0.48\textwidth]{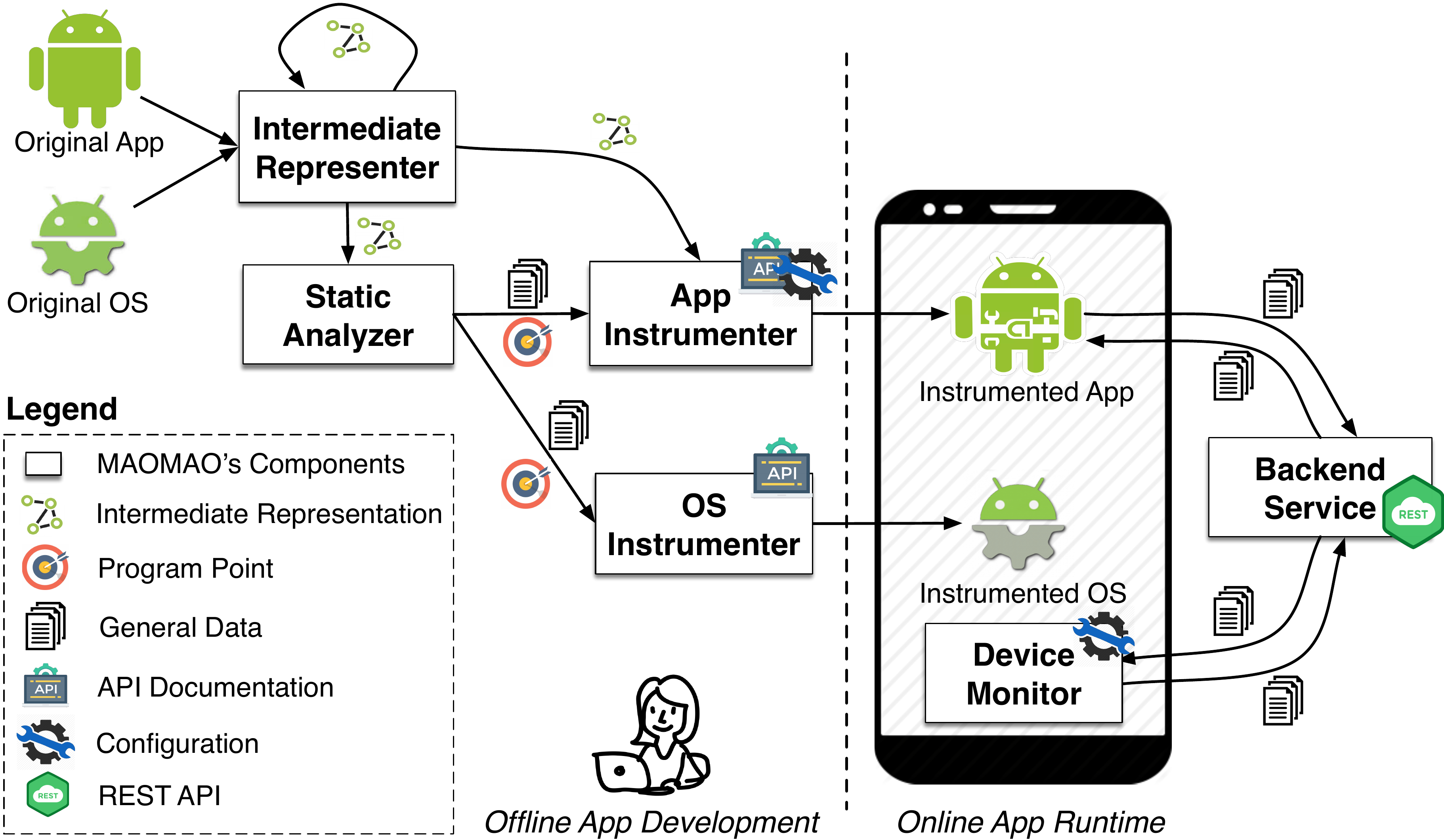}
        \vspace{-2mm}
	\caption{\appr's six reference components and overall workflow} 
		\vspace{-5mm}
	\label{fig:maomao}
\end{figure}

As Fig.~\ref{fig:maomao} shows, \appr's reference architecture consists of six components, i.e., microservices.
An individual app optimization technique can consist of one or more of the reference components. 
For example, {IccTA}~\cite{li2015iccta} only has the \textit{Intermediate Representer} and \textit{Static Analyzer}. 

\textit{\underline{Intermediate Representer}} takes an app or the Operating System (OS), e.g., Android framework, as the input and produces an Intermediate Representation (IR) for \textit{Static Analyzer} to analyze. IR can be used by other Intermediate Representer to build new IR. For example, tool-specific IR is usually built on top of fundamental IRs, such as Abstract Syntax Tree (AST), Control Flow Graph (CFG) of an app. {GATOR}~\cite{yang2015ccfg} has an \textit{Intermediate Representer} to produce Callback Control Flow Graph (CCFG), which is a tool-specific IR that uses CFG.

\textit{\underline{Static Analyzer}} analyzes the IR to extract useful information that can be used in other components, such as the program point to be instrumented that will be used to instrument the app or the OS. For instance, {PerfChecker}~\cite{liu2014characterizing} has a \textit{Static Analyzer} to detect performance bugs, which can be used by app developers directly or reused by follow-up techniques to fix the bugs based on the bug locations (i.e., program point). 

\textit{\underline{App Instrumenter}} instruments the app code and transforms the original app, usually based on the information extracted from the \textit{Static Analyzer}. The \textit{App Instrumenter} can be categorized into \textit{Automatic App Instrumenter} (AAI) or \textit{Manual App  Instrumenter} (MAI), and it usually needs to be configured so that the instrumented app can interact with other specific components at runtime, such as \textit{Backend Service}. An AAI instruments the app  without developer's involvement, e.g., {PALOMA}'s~\cite{paloma_icse} \textit{Instrumenter} is an AAI used to enable prefetching based on the information extracted from {PALOMA}'s \textit{Static Analyzer}s.
On the other hand, a MAI provides APIs for developers to manually modify their code. 

\textit{\underline{OS Instrumenter}} is similar to \textit{App Instrumenter}, but it instruments the OS (e.g., Android) instead of the app. \textit{OS Instrumenter}s can also be categorized into \textit{Automatic OS Instrumenter}s (AOSI) and \textit{Manual OS Instrumenter}s (MOSI). For instance, {SEALANT}'s \textit{Interceptor} is a MOSI that extends the Android framework to block malicious intents at runtime. 

\textit{\underline{Device Monitor}} observes the device-level conditions at app runtime. It is typically used to balance the quality-of-service (QoS) trade-offs since mobile devices are resource-constrained. Similar to the \textit{App Instrumenter}, it also needs to be configured in order to interact with other components at runtime, such as the \textit{Backend Service}. For instance, {IMP}'s \textit{Monitor}  is a \textit{Device Monitor} targeting battery life, data usage, and network bandwidth, that interacts with its \textit{Prefetcher}.

\textit{\underline{Backend Service}} contains the ancillary  functionalities that are triggered at app runtime. It interacts with the instrumented app and the \textit{Device Monitor} via lightweight protocol, e.g., REST APIs~\cite{fielding2000rest}. The ancillary functionalities are usually triggered by specific information sent from the instrumented app or the \textit{Device Monitor}. For instance, {IMP}'s \textit{Prefetcher}  is a \textit{Backend Service} that adapts its prefetching strategies according to the device's QoS conditions sent by its \textit{Monitor}.

Fig.~\ref{fig:ra-api} shows the reference APIs for each reference component that aims to aid the design and implementation of \appr, and a concrete example will be shown in Section~\ref{sec:maomao:implement}. 

\subsection{Migration of Existing Techniques to \appr}
\label{sec:maomao:implement}

\begin{figure}
       \vspace{-5mm}
	\centering
	\includegraphics[width=0.4\textwidth]{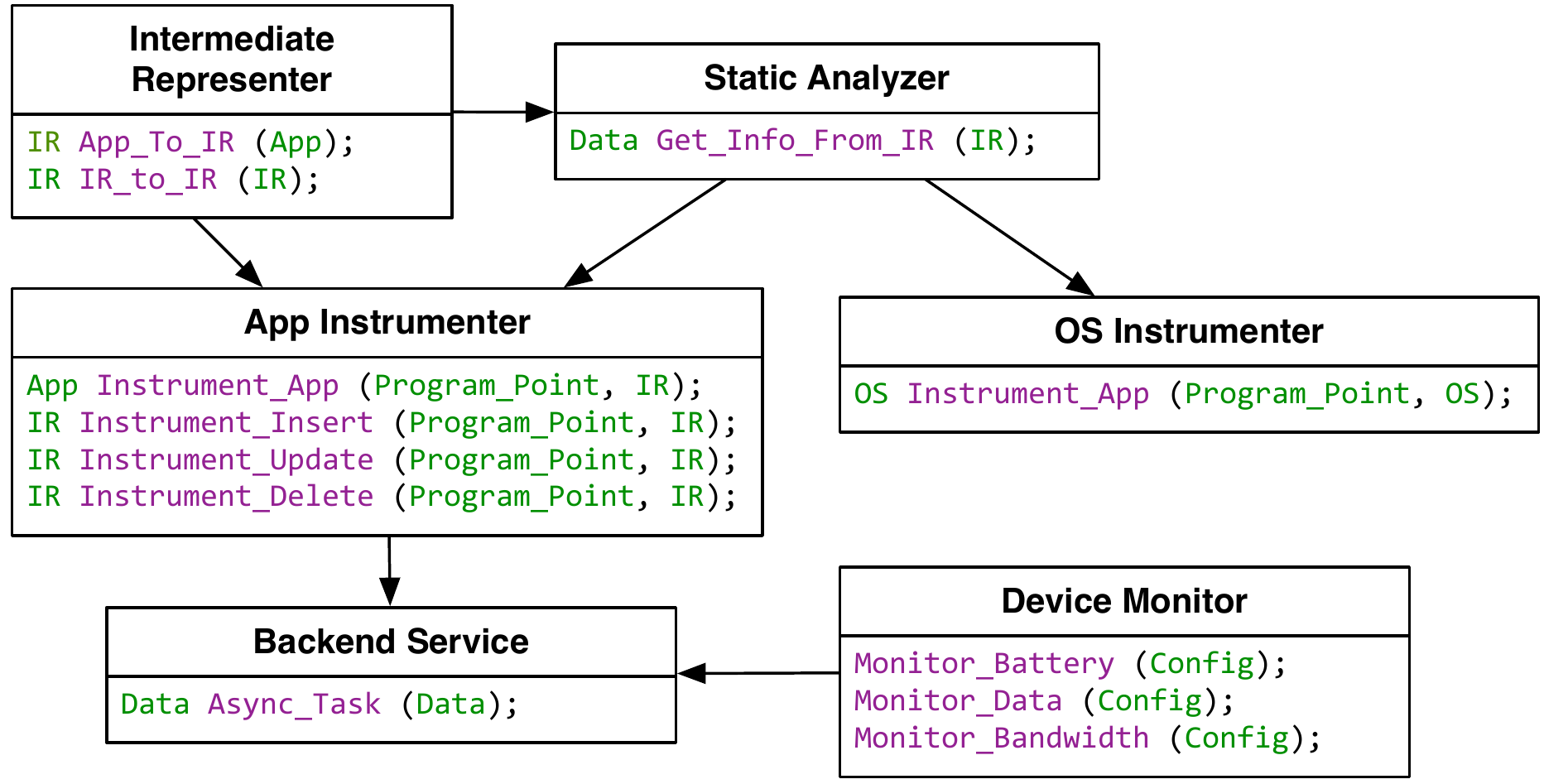}
        \vspace{-3mm}
	\caption{\appr's reference components and their reference APIs}
		\vspace{-5mm}
	\label{fig:ra-api}
\end{figure}


We hypothesize that designing new techniques using \appr's microservice architecture will be more straightforward than migrating an existing technique. For this reason, in this section we illustrate how the latter process can be approached
using examples from Sections~\ref{sec:example}~and~\ref{sec:maomao:design}. Specifically, we studied the designs of the existing techniques as well as their available open-source implementations to establish that they can be ported to \appr's architecture. 

\begin{figure}
	\vspace{-5mm}
	\centering
		\includegraphics[width=0.4\textwidth]{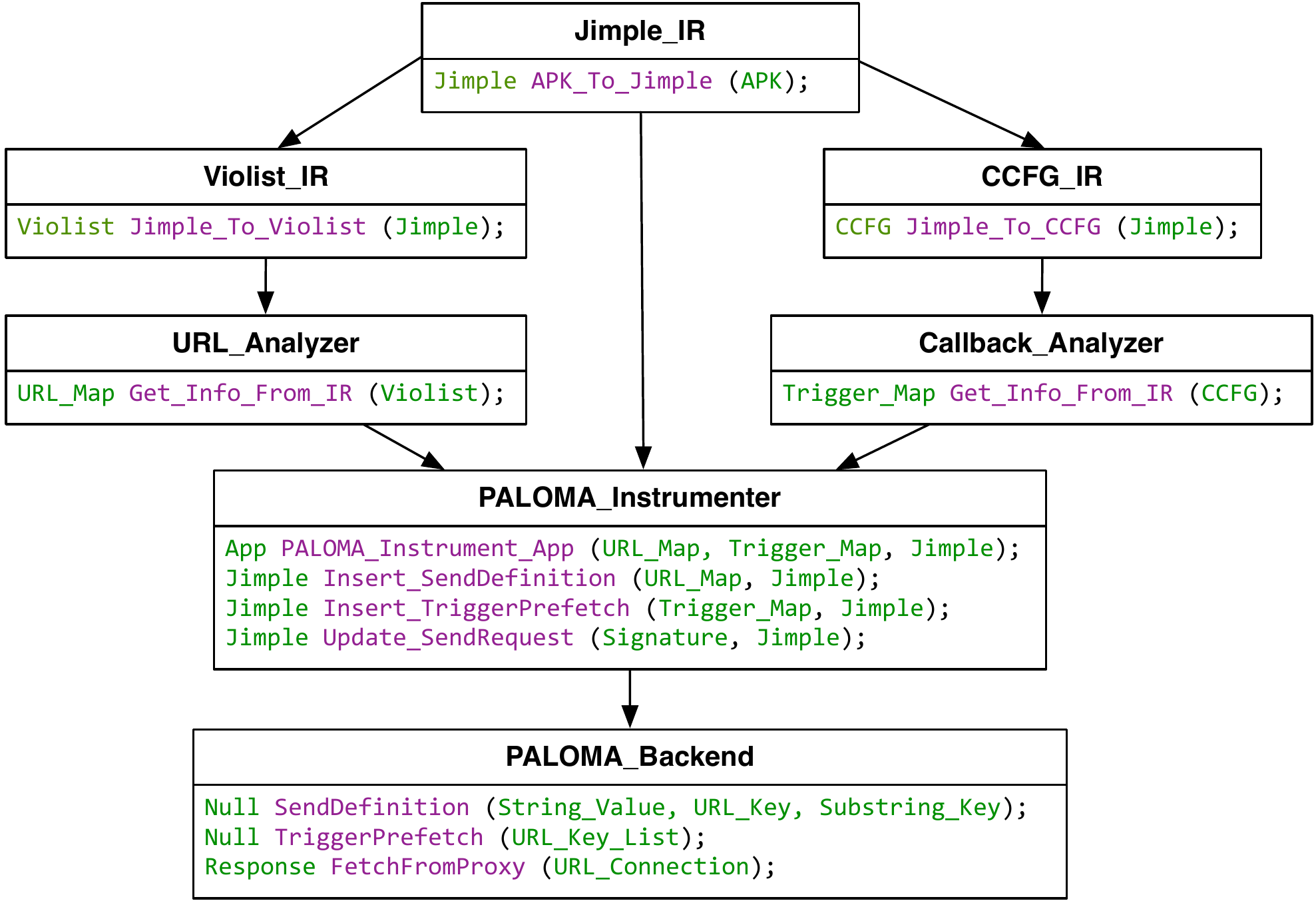}
        \vspace{-2mm}
	\caption{The class diagram of PALOMA in \appr} 
		\vspace{-5mm}
	\label{fig:paloma-class}
\end{figure}

Table~\ref{tbl:mapping} shows the mapping between the components in the existing techniques and \appr's components. We use {PALOMA}~\cite{paloma_icse} as an example to explain the details of the mapping. Fig.~\ref{fig:paloma-class} shows the class diagram of {PALOMA} when migrated to \appr~following the reference components and APIs shown in Fig.~\ref{fig:ra-api}. {PALOMA} was selected because it was recently published and it contains more components to be migrated to \appr's architecture than the other techniques, which have as few as a single relevant component.

\begin{table}[b!]
	\vspace{-5mm}
\centering
\caption{Mappings between Existing Components in Existing Techniques (Section~\ref{sec:example}) and \appr's Components (Section~\ref{sec:maomao:design})}
	\vspace{-3mm}
\label{tbl:mapping}
\centering
\resizebox{\linewidth}{!}{
\begin{tabular}{|c|c|c|}
\hline
\textbf{Existing Technique} & \textbf{Existing Component} & \textbf{MAOMAO's Component}                             \\ \hline
\multirow{4}{*}{PALOMA~\cite{paloma_icse}}     & \textit{String Analyzer}    & Intermediate Representer + Static Analyzer              \\ \cline{2-3} 
                            & \textit{Callback Analyzer}  & Intermediate Representer + Static Analyzer              \\ \cline{2-3} 
                            & \textit{Instrumenter}   & Intermediate Representer + (Automatic) App Instrumenter \\ \cline{2-3} 
                            & \textit{Proxy}              & Backend Service                                         \\ \hline
\multirow{3}{*}{IMP~\cite{higgins2012informed}}        & \textit{API Support}        & (Manual) App Instrumenter                               \\ \cline{2-3} 
                            & \textit{Monitor}            & Device Monitor                                          \\ \cline{2-3} 
                            & \textit{Prefetcher}         & Backend Service                                         \\ \hline
\multirow{3}{*}{Bouquet~\cite{li16bouquet}}    & \textit{Detector}           & Intermediate Representer + Static Analyzer              \\ \cline{2-3} 
                            & \textit{Bundling Analyzer}  & Static Analyzer + (Automatic) App Instrumenter          \\ \cline{2-3} 
                            & \textit{Proxy}              & Backend Service                                         \\ \hline
IccTA~\cite{li2015iccta}                       & \textit{Taint Analyzer}     & Intermediate Representer + Static Analyzer              \\ \hline
\multirow{2}{*}{SEALANT~\cite{lee2017sealant}}    & \textit{Analyzer}           & Intermediate Representer + Static Analyzer              \\ \cline{2-3} 
                            & \textit{Interceptor}        & (Manual) OS Instrumenter                                \\ \hline
ApkCombiner~\cite{li2015apkcombiner}                 & \textit{Combiner}           & (Automatic) App Instrumenter                            \\ \hline
\end{tabular}
}
\end{table}

\paloma's \textit{String Analyzer} leverages an external string analysis tool, 
{Violist}~\cite{li2015string}, to identify the string values of URLs in an app, and outputs a \texttt{URL Map}. {Violist} has a proprietary intermediate representation (IR) of the control- and data-flow relationships among the string variables and string operations. 
Violist's IR is transformed from Jimple~\cite{bartel2012dexpler}, which is a fundamental IR for representing Java/Android programs. 
{Violist}  analyzes the string values at given program points based on its IR. With \appr, \paloma's \textit{String Analyzer} will be implemented as two \emph{Intermediate Representer} microservices (\texttt{Jimple\_IR},  \texttt{Violist\_IR}) and one \emph{Static Analyzer} microservice (\texttt{URL\_Analyzer}) as shown in Fig.~\ref{fig:paloma-class}. \texttt{Jimple\_IR}'s \texttt{APK\_To\_Jimple} API is an implementation of \emph{Intermediate Representer}'s \texttt{App\_To\_IR} (Fig.~\ref{fig:ra-api}), where \texttt{Jimple} is an instance of \texttt{IR}, and \texttt{APK}~\cite{apk}
is an instance of \texttt{App}. Similarly, \texttt{Violist\_IR}'s \texttt{Jimple\_To\_Violist}  is an implementation of \emph{Intermediate Representer}'s \texttt{IR\_To\_IR}, where \texttt{Violist} and \texttt{Jimple} are both instances of \texttt{IR}. \texttt{URL\_Analyzer}'s \texttt{Get\_URLMap\_From\_Violist} is an implementation of \emph{Static Analyzer}'s \texttt{Get\_Info\_From\_IR}.

\paloma's \textit{Callback Analyzer} leverages an external callback analysis tool, {GATOR}~\cite{yang2015ccfg}, to identify the data prefetching points in an app, and generates a \texttt{Trigger Map}. Specifically, the \textit{Callback Analyzer} relies on the CCFG defined by {GATOR}. With \appr, the \textit{Callback Analyzer} is decomposed into two microservices:  an \textit{Intermediate Representer} (\texttt{CCFG\_IR}) that outputs the CCFG by reusing \texttt{Jimple\_IR}, and a \textit{Static Analyzer} (\texttt{Callback\_Analyzer}) that outputs the \texttt{Trigger\_Map} for instrumenting prefetching functions at given program points based on the CCFG.

\paloma's \textit{Instrumenter} takes as inputs the \texttt{URL\_Map}, the \texttt{Trigger\_Map}, and the \texttt{Jimple}, and transforms the original app to a prefetching-enabled app with three instrumentation functions. With \appr, \paloma's \textit{Instrumenter} will be one \emph{App Instrumenter} (\texttt{\paloma\_\allowbreak Instrumenter}) that instruments the app based on the outputs from \texttt{URL\_Analyzer}, \texttt{Callback\_Analyzer}, and  \texttt{Jimple\_IR} as shown in Fig.~\ref{fig:paloma-class}. 

Finally, \paloma's \textit{Proxy} interacts with the instrumented app at runtime via the three instrumented functions: (1)~it updates the \texttt{URL Map} with the data sent by \texttt{SendDefinition}; (2)~it prefetches HTTP requests triggered by  \texttt{TriggerPrefetch}; and (3)~it redirects the HTTP requests  to get the response from a cache triggered by \texttt{FetchFromProxy}. Migrating  \paloma's \textit{Proxy} to \appr~ is straightforward: it will be designed as a single \textit{Backend Service} microservice (\texttt{\paloma\_Backend}), with three REST APIs to represent the three instrumented functions.  

Other existing mobile computing techniques would be redesigned (and subsequently reimplemented) in an analogous fashion. As mentioned previously, new techniques would follow \appr's reference architecture and rely on its APIs from the get-go. 

\section{\appr's envisioned adoption}
\label{sec:vision}


\appr's architecture allows app optimization techniques to be decomposed into lightweight reusable  microservices with a standard workflow, enabling their use by  both developers and researchers. Specifically, \appr~alleviates
the problem of reusing often incompatible capabilities from disparate research techniques.

To realize \appr's potential in practice, we propose a \textit{Microservice Repository} (MR), that aims to connect  developers and researchers together by providing and enforcing a shared baseline. 
MR is a cloud-based  repository that consists of a \textit{Service Request Pool} (SRP) and a \textit{Microservice Pool} (MP). SRP stores developers' requests for their desired services. MP stores and provides access to the available microservices  and their corresponding API documentation. 

We use mobile app security techniques---{IccTA}~\cite{li2015iccta}, {SEALANT}~\cite{lee2017sealant}, and {APKCombiner}~\cite{li2015apkcombiner}---to demonstrate how developers and researchers can benefit from \appr~and MR. We choose  security because it has attracted the greatest  attention among researchers in the mobile domain. 

As Table~\ref{tbl:mapping} shows, {IccTA}'s \textit{Taint Analyzer} is decompsosed into \textit{Intermediate Representer} (IR)  and \textit{Static Analyzer} (SA) microservices in \appr's architecture. Similarly, {SEALANT} consists of IR, SA, and \textit{Manual OS Instrumenter} (MOSI) microservices. Finally, {ApkCombiner}  becomes an \textit{Automated App Instrumenter} (AAI) microservice. 

The six \appr~microservices in the three techniques will be deployed to MR's MP, along with their corresponding API documentation. 
We discuss several representative use cases of developers and researchers using these services. 

\begin{enumerate}
    \item Both developers and researchers can search the MR to find their desired microservices in a specific domain  (e.g., mobile security domain).
    \item {IccTA}'s authors can find {ApkCombiner}'s AAI in the MR and extend {IccTA}'s SA to detect \textit{inter-app} privacy leaks by following the API documentation of {ApkCombiner}'s AAI. Then, {IccTA}'s optimized inter-app SA can be deployed as a new microservice to the MR. 
    \item {SEALANT}'s authors can extend its SA in the same manner as {IccTA}. {SEALANT} and {IccTA} can then use each other's IR and SA microservices to  compare the two solutions. Moreover, since {SEALANT}'s MOSI outputs an instrumented OS to block privacy leaks, it can be used by {IccTA}  to ``upgrade''  from detection to optimization.
    \item A phone manufacturer's engineers  can find {SEALANT}'s MOSI in the MP and follow its  APIs to customize the OS to block privacy leak  on their phones.
    \item If developers cannot find a desired microservice in the MR, they can submit a service request to the SRP to describe their needs 
    (e.g., a request for a performance bottleneck detection service). They can optionally attach a benchmark app that has the relevant  issue. 
   \item Researchers can search the SRP to find reported needs in their  domain of interest and possibly obtain the corresponding testing data (e.g., using submitted benchmark apps to evaluate a performance bug detection technique). 
\end{enumerate}

\appr's microservices and MR allow researchers to track the real-world needs and  developers to adopt research techniques readily by invoking lightweight APIs. An added advantage is that  the microservices are deployed on the cloud and do not introduce significant overhead on the client apps deployed on resource-constrained mobile devices. Researchers can also dynamically update their microservices without requiring modifications to the app code. In addition, the testing data provided by developers in the SRP can serve as benchmarks for comparing different techniques in the same domain. Once the microservices are adopted by developers, the underlying research techniques will be evaluated in the real world with real users, providing insights and incentives for researchers to improve their techniques. 


\section{Conclusion and Future Work}
\label{sec:conclusion}

We introduce \appr, a reference architecture for mobile app optimization techniques to guide the design of future techniques in order to  improve their reusability and extensibility, with a corresponding {Microservice Repository} (MR) to deploy \appr-compatible techniques. Together, the two improve the availability and practicality of research techniques, and bridge the gap between researchers and developers. Our preliminary work provides evidence of \appr's and MR's viability, and also shows several future  research directions in order to adopt \appr~ in practice,
such as ensuring privacy of any data (e.g., app usage) submitted to MR, scalability to large numbers of researchers and developers, and standardizing API documentations. 
As early versions of \appr~and MR are deployed and adopted, this scope will grow to include capabilities such as recommenders of related microservices based on certain service requests and an access control model to enable fine-grained data sharing.

\clearpage
\bibliographystyle{IEEEtran}
\bibliography{MOBILESoft2019} 

\end{document}